\documentclass[conference,a4paper]{IEEEtran}

\usepackage[T1]{fontenc}
\usepackage{mathptmx}  
\usepackage[scaled=.92]{helvet}  
\usepackage{courier}  

\IEEEoverridecommandlockouts
\usepackage{cite}

\ifCLASSINFOpdf
  \usepackage[pdftex]{graphicx}
  \graphicspath{{../pdf/}{../jpeg/}{../figures/}}
\else
\fi
\ifCLASSOPTIONcompsoc
\usepackage[caption=false, font=normalsize, labelfont=sf, textfont=sf]{subfig}
\else
\usepackage[caption=false, font=footnotesize]{subfig}
\usepackage{booktabs}
\usepackage{multirow}
\usepackage{color}

\usepackage{amsmath}
\usepackage{amsthm}
\usepackage{algorithm}
\usepackage{algorithmic}

\usepackage{array}

\ifCLASSOPTIONcompsoc
 \usepackage[caption=false,font=normalsize,labelfont=sf,textfont=sf]{subfig}
\else
 \usepackage[caption=false,font=footnotesize]{subfig}
\fi

\usepackage{fixltx2e}
\usepackage{dblfloatfix}

\ifCLASSOPTIONcaptionsoff
 \usepackage[nomarkers]{endfloat}
\let\MYoriglatexcaption\caption
\renewcommand{\caption}[2][\relax]{\MYoriglatexcaption[#2]{#2}}
\fi

\usepackage{url}

\begin{document}
%
\title{Integrating Fast-response Capability into Virtual Power Plant Operation for Ancillary Services
}
%
%
%

\author{\IEEEauthorblockN{Qixing Liu$^{1}$, Ruike Lyu$^{2}$, Zhe Zhai$^{1}$, Yan Shen$^{2}$, Xue Liu$^{2}$, and Hongye Guo$^{*2}$}
  \IEEEauthorblockA{$^{1}$Electric Power Dispatching and Control Center, China Southern Power Grid, Guangzhou, China}
  \IEEEauthorblockA{$^{2}$State Key Laboratory of Power System Operation and Control, Dept. of Electrical Eng., Tsinghua Univ., Beijing, China}

  \thanks{Corresponding author: Hongye Guo (Email: hyguo@tsinghua.edu.cn). This work was supported by THU-CSPG Digital Power Grid Joint Research Institute Project (WT0000550000001088).}

}

\maketitle

\begin{abstract}
Virtual power plants (VPPs) can aggregate distributed energy resources (DERs) to provide ancillary services for power systems, creating new profit opportunities. Ancillary services such as secondary frequency regulation require providers to have sufficient response capability to follow rapidly changing control commands. If overlooking the response requirement, the VPP will not be able to accurately measure its regulation capability, reducing its earnings in performance-based markets or risking disqualification. This paper integrates the requirement for fast-response capability into the operational framework of VPPs providing ancillary services. We leverage historical control commands to formulate chance constraints in the bidding model, mandating that the VPP's fast-response capability meets the requirement of ancillary services with a specified probability. Case studies verify that considering fast-response capabilities can enhance VPP operation.
\end{abstract}

\begin{IEEEkeywords}
Virtual power plant, ancillary services, fast-response, chance constraint, bidding strategy.
\end{IEEEkeywords}

\ifCLASSOPTIONpeerreview
  \begin{center} \bfseries EDICS Category: 3-BBND \end{center}
\fi
%
\IEEEpeerreviewmaketitle





\section{Introduction}
%
%
%
%
\IEEEPARstart{V}{irtual} power plants (VPPs) represent a crucial technological approach for effectively managing the burgeoning multitude of distributed energy resources (DERs)~\cite{yi_aggregate_2021}. In the low-carbon transition of energy systems, the integration of a high proportion of fluctuating renewable energy sources, such as photovoltaic and wind power, presents unprecedented challenges to the real-time balance of power systems and increases the demand for ancillary services such as frequency regulation~\cite{kroposki_achieving_2017}. On the load side, DERs such as electric vehicles (EVs) and customer-side energy storage (ES) are small in size but large in number, making them challenging for grid operators to directly monitor and control~\cite{zhou2020optimal}. By utilizing advanced communication and control methods, VPPs can effectively monitor and manage DERs, aggregating them to exhibit controllable external characteristics akin to traditional power plants and providing frequency regulation to support the operation of power systems~\cite{golpira_enhanced_2024, dallanese_optimal_2018, ruan_data-driven_2024, naughton2021co}.

Typical ancillary services, such as frequency regulation, place high demands on the performance of providers. Most existing markets are designed to be performance-based~\cite{he_optimal_2016}, meaning that the performance in following control commands will determine the eligibility and market revenue. Given the randomness and rapid changes in control commands, the technical threshold required to provide ancillary services can be relatively high for DERs. Therefore, the commercial viability of a VPP largely depends on its ability to help DERs meet the threshold for ancillary service markets and generate surplus revenue through advanced forecasting and decision-making~\cite{ruan_data-driven_2024}.
For market bidding, the VPP organizes the energy usage of DERs and determines energy purchase and ancillary service bids to maximize profits~\cite{naughton2021co}. When deploying ancillary services, the VPP manages the output powers of DERs such that they collectively respond to control commands for ancillary services~\cite{dallanese_optimal_2018}.
Advanced operation creates additional profitability compared to individual DERs and is the basis for formulating VPPs~\cite{chen_competition_2022}.

\begin{figure}[!t]
  \centering
  \includegraphics[width=3.49in]{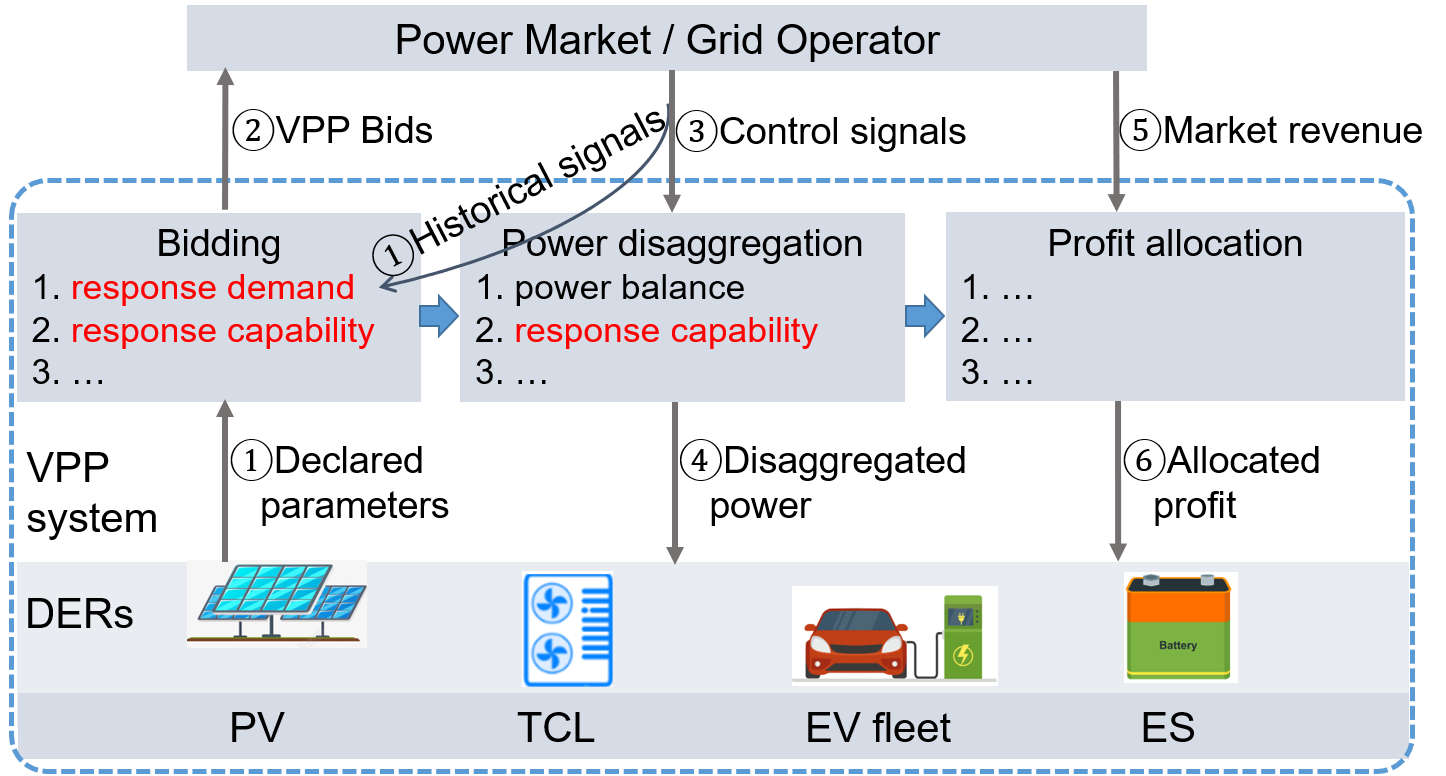}
\caption{To incorporate considerations of fast-response capability, modifications were made to the VPP operation framework in this paper (marked in red).}
  \label{fig_framework}
\end{figure}

However, in the existing research on the provision of ancillary services by VPPs, fast-response capabilities have not been adequately considered. Specifically, this paper focuses on the fast-response capability required by a VPP when responding to real-time control commands for ancillary services such as frequency regulation within the dispatch interval~\cite{article1}. This refers to the maximum change in power per unit time at the second-to-minute timescale. Power electronic resources such as electrochemical energy storage and electric vehicles can quickly change the power they exchange with the grid. Therefore, a VPP primarily composed of these resources only needs to ensure that its power and energy capacity meet the relevant requirements~\cite{chen_real-time_2024}. However, the range of resources aggregated by the VPP is expanding to include large power capacity loads with limited adjusting capabilities, such as thermostatically controlled loads (TCLs)~\cite{cui_evaluation_2018}. In this context, if the fast-response capabilities of DERs are overlooked, the resulting VPP bidding and power control strategies may prove unfeasible when responding to grid commands, seriously damaging the ability of VPPs to generate surplus revenue~\cite{en14134066}.

In order to incorporate fast-response capabilities into VPP operations, a primary challenge lies in modeling the requirements for fast response in ancillary services. The regulation signal and the reserve signal are time series sent in real time by the grid operator (GO) to resources providing ancillary services through an automatic closed-loop feedback control system to respond to events such as load fluctuations and unit failures. From the perspective of the VPP, the parameters and operating state of the control system are unknown, and events such as load fluctuations and unit failures are highly random. Therefore, before the VPP actually receives the control signal (for example, when considering the feasibility of frequency regulation capacity in bidding), it can only model the signals as a random sequence and use historical signals to estimate the distribution.

In this paper, we integrate the consideration of resource fast-response capabilities into the operational framework of VPPs providing ancillary services, which includes the stages of bidding and power allocation (Fig.\ref{fig_framework}). Specifically, we use historical data to model the probabilistic fast-response demand of ancillary services and embed it into the VPP bidding model as a chance constraint. This consideration takes into account the cooperation of heterogeneous resources to meet the demand. Our method can avoid overly diminishing the VPP's performance score due to insufficient fast-response capabilities, contributing to enhancing the technical and commercial viability of VPPs to exploit the flexibility of DERs.

\section{Ancillary Service Market Framework and Control Signal Model}\label{model_market}

\subsection{Ancillary Service Market Framework}

We consider two typical ancillary services: regulation (secondary frequency regulation) and reserve (spinning reserve). As different markets have different names and definitions for ancillary services, we adopt the general market model for exposition.
Before actually providing ancillary services, the VPP needs to submit its bids to the market. Considering the relatively small capacity of the VPP, we regard it as a price-taker that performs self-dispatch, with the bid form being hourly energy output $\tilde{p}_t$, the capacity to provide regulation and reserve $Cap^{\rm reg}_t$ and $Cap^{\rm res}_t$, respectively, and $t = 1,..., T$, where $T$ is the ending time slot of the said horizon. After the market is cleared, the VPP receives its accepted capacity, which should be consistent with the capacity it has submitted since price takers can declare the desired capacities at a zero price to ensure clearance.

When actually providing ancillary services, the GO sends control signals to the VPP based on the system state, including the normalized regulation signal (AGC signal) $\delta^{\rm reg}_{\hat{t}}$ and the reserve signal $\delta^{\rm res}_{\hat{t}}$, where $\hat{t}$ is the index for the time interval of the control signals. The overall adjustment command is scaled by the capacity, that is, $\delta^{\rm reg}_{\hat{t}} Cap^{\rm reg}_t + \delta^{\rm res}_{\hat{t}} Cap^{\rm res}_t$. The VPP needs to adjust the overall output based on the baseline output (determined by bid $\tilde{p}_t$ in the energy market) to follow the received control command, thereby satisfying the GO's assessment. Naturally, after deviating from the baseline output, the energy exchanged between the VPP and the grid deviates from the energy bid. The deviated energy is usually settled at the energy price to avoid duplicate compensation for ancillary services.

\subsection{Control Signal Model}

Taking the provision of secondary frequency regulation as an example, the modeling of the control signal mainly has the following three key factors:

\paragraph{Distribution of individual signals} First, ignoring the temporality of the signals, each individual regulation signal is regarded as a random variable, and its probability space is modeled using historical data. The sample space of the regulation signal is $\delta^{\rm reg} \in \Delta^{\rm reg}=[-1, 1]$. For convenience of numerical calculation, the sample space is discretized to obtain the event space, for example, $S^{\rm reg}=\{\{-1\}, (-1, -0.9], ..., [0.9, 1), \{1\}\}$. Following the convention of stochastic optimization, we refer to an event in the set of signal events as a scenario. The probability of scenario $s$, i.e., the regulation signal $\delta^{\rm reg}$ belonging to the interval $s \in S^{\rm reg}$, can be estimated using historical signals.

\paragraph{Worst-case scenario for the signals} The actual signal may deviate from the estimated distribution. To address this, distribution constraints or robust constraints can be added to ensure that the VPP can follow commands under the worst-case control signal scenarios (for example, when the GO continues to require resources to increase power injection into the grid at maximum capacity). In fact, to ensure system safety, the current market mechanism usually requires the ability of resources to maintain maximum output.
For example, PJM requires resources providing regulation to be able to maintain maximum regulation power for at least 15 minutes. The market requirements can be incorporated into the bidding model, for example, by adding the following constraints for the ES unit:
\begin{subequations}\label{model_req_reg}
  \begin{equation}\label{model_req_reg1}
    e_{t} - \underline{e} \ge \frac{1}{\eta^{\rm dis}} \Delta t^{\rm req} (\tilde{p}_t + Cap^{\rm reg}_t)
  \end{equation}
  \begin{equation}\label{model_req_reg2}
    \overline{e} - e_{t} \ge {\eta^{\rm ch}} \Delta t^{\rm req} (- \tilde{p}_t + Cap^{\rm reg}_t)
  \end{equation}
\end{subequations}
where $\eta^{\rm dis}$ and $\eta^{\rm ch}$ are the charging and discharging efficiency of the resource, $\Delta t^{\rm req}$ is the required time for the resource to maintain the maximum output, and $\underline{e}_{t}$ and $\overline{e}_{t}$ are the lower and upper bounds of the energy state of the resource, respectively. The market requirements are usually higher than the actual needs, so after incorporating the requirements in the operation strategy, it is not necessary to specifically model the worst-case scenario for the signals.

\begin{figure}[!t]
  \centering
  \includegraphics[width=3.0in]{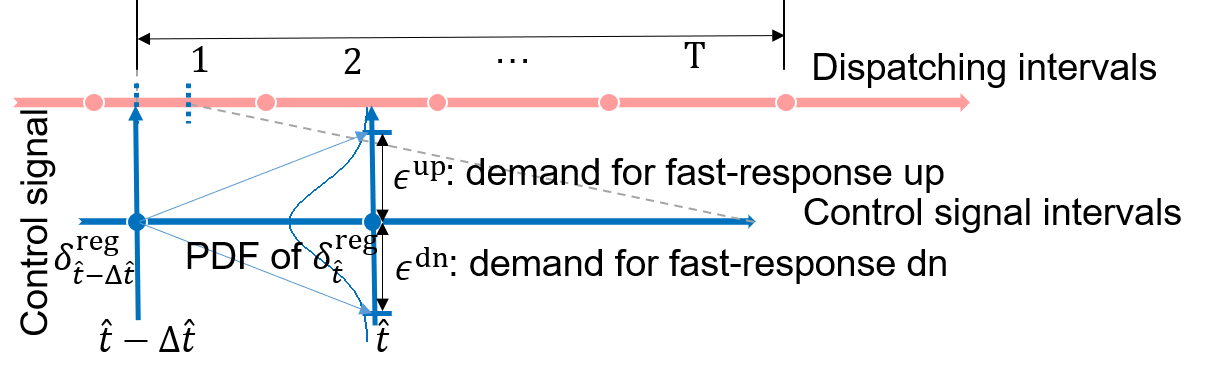}
\caption{The demand for the fast-response capability of ancillary services is represented by using the quantiles of the historical control signal ramp rate distribution.}
  \label{fig_framework_time}
\end{figure}

\paragraph{Fast-response demand} To ensure system safety, ancillary services represented by frequency regulation require resources to respond quickly to control signals, which may change from the minimum to the maximum value within a minute. If the response capability of the resource is insufficient, it will lead to poor response speed and result in a low performance score, thereby reducing the resource's revenue or even canceling its qualifications. To model the response requirements in most cases, the ramp rate between two signals is defined as the random variable $\varepsilon = (\delta^{\rm reg}_{\hat{t}} - \delta^{\rm reg}_{\hat{t} - \Delta \hat{t}})/\Delta \hat{t}$, $\Delta \hat{t}$ is the length of the sampled control signal intervals, and the response demand is determined by the quantiles of $\varepsilon$ (Fig.~\ref{fig_framework_time}):
\begin{equation}\label{model_signal_ramp}
  \begin{aligned}
     & \left\{\begin{array}{c}
      R_t^{\rm up}=\varepsilon^{\rm up}_t \cdot Cap_t^{\rm reg} \\
      \int_{-\infty}^{\varepsilon^{\rm up}_t} p_t(\varepsilon) {\rm d} \varepsilon= c^{\rm up}
    \end{array}\right.  \\
     & \left\{\begin{array}{c}
      R_t^{\rm dn}= - \varepsilon^{\rm dn}_t \cdot Cap_t^{\rm reg} \\
      \int_{-\infty}^{\varepsilon^{\rm dn}_t} p_t(\varepsilon) {\rm d} \varepsilon= c^{\rm dn}
    \end{array}\right.
  \end{aligned}
\end{equation}
where $R_t^{\rm up(dn)}$ is the up(down)-response demand, $\varepsilon^{\rm up(dn)}$ is the $c^{\rm up(dn)}$ quantile of the ramp rate, $c^{\rm up} - c^{\rm dn}$ is the confidence level, and $p_t(\varepsilon)$ is the probability density function of the ramp rate, which can be estimated using historical data. During bidding, the fast-response capability can be adjusted by changing the value of $c^{\rm up/dn}$, thereby adjusting the performance score. For example, setting $c^{\rm up}=0.95$ and $c^{\rm dn}=0.05$ means that in 90\% of the cases, the resource's fast-response capability allows it to fully follow the control signal, which should result in a good performance score.

\section{Fast-response Capability Constraints for VPP Bidding}\label{sec_model_operation}

Due to space constraints, we only present the modifications made to the state-of-the-art VPP operational model. Unaltered sections, such as the VPP profit model and other DER operation constraints, are assumed to follow the model described in Reference \cite{chen_real-time_2024} (see Appendix~\ref{apd_model}). 
In the subsequent mathematical formulation, we use subscripts $t, s, i$ to denote variables pertinent to the time interval $t$, GO signal scenario $s$, and DER $i$. We use subscripts ${\rm ch(dis)}$ and ${\rm init}$ to represent the (dis)charging and initial states, respectively. For instance, $p^{\rm ch}_{t, s, i}$ represents the charging power of resource $i$ during time interval $t$ under scenario $s$, with ``charging'' implying power drawn from the grid. In terms of parameter notation, we use underlines/overlines to indicate lower/upper boundaries. The state of resources, e.g., the battery state of charge (SOC) or indoor temperature, is denoted by $e$. The state of DER $i$ at the onset of time interval $t$ is represented by $e_{t, i}$; let $e_{T+1, i}$ represent the state after the time horizon, where $T$ is the ending time slot. $p^{\rm dis}_{t, s, i}$ and $p^{\rm ch}_{t, s, i}$ are the discharging and charging power of DER $i$ at time interval $t$ and signal scenario $s$.

\paragraph{Capacity constraints} We define the up (dn) response capacity of DER $i$ as the range by which it can deviate upward (downward) from the baseline output, expressed as:
\begin{equation}\label{model_def_capacity}
  Cap_{t, i}^{\rm up}  := \underset{s}{\rm max.}\{p_{t, s, i}\} - \tilde{p}_{t, i}, \ Cap_{t, i}^{\rm dn} := \tilde{p}_{t, i} - \underset{s}{\rm min.}\{p_{t, s, i}\}
\end{equation}
where $Cap_{t, i}^{\rm up}$ and $Cap_{t, i}^{\rm dn}$ are the up-response and down-response capacities provided by resource $i$, respectively, and $\tilde{p}_{t, i}$ is the baseline power of DER $i$, i.e., the allocated power in the signal scenario of $(\delta^{\rm reg}_s, \delta^{\rm res}_s)=(0, 0)$.
$\underset{s}{\rm max.}\{p_{t, s, i}\}$ and $\underset{s}{\rm min.}\{p_{t, s, i}\}$ are the maximum and minimum powers of DER $i$ in the signal scenarios, respectively.
The VPP needs to ensure that the total response capacity of the resources satisfies the bids for regulation and reserve services, formulated as ($\forall t$):
\begin{subequations}\label{model_capacity}
  \begin{equation}
    \sum_i Cap_{t, i}^{\rm up}  \ge Cap_t^{\rm reg}+Cap_t^{\rm res}, \sum_i Cap_{t, i}^{\rm dn} \ge Cap_t^{\rm reg}
  \end{equation}
  \begin{equation}
    (Cap_{t, i}^{\rm up}, Cap_{t, i}^{\rm dn}, Cap_t^{\rm reg}, Cap_t^{\rm res}) \ge 0.
  \end{equation}
\end{subequations}

\paragraph{Power balance} In each signal scenario, the VPP needs to ensure that the total power output of the resources matches the control signal, formulated as ($\forall t \forall s$):
\begin{equation}
  \sum_i (p^{\rm dis}_{t, s, i} - p^{\rm ch}_{t, s, i}) = \tilde{p}_t + \delta^{\rm reg}_s Cap^{\rm reg}_t + \delta^{\rm res}_s Cap^{\rm res}_t, \ \forall t \forall s.
\end{equation}

\paragraph{Energy reserve ratio} As mentioned above, the VPP needs to ensure that the energy reserve ratio of the resources satisfies the requirements of the ancillary service market, formulated as ($\forall t \forall s \forall i$):
\begin{equation}\label{model_energy_reserve}
  \underline{e}_{t, i} \le \theta_i e_{t, i} + (\eta^{\rm ch}_{i} p^{\rm ch}_{t, s, i} -  \frac{1}{\eta^{\rm dis}_{i}} p^{\rm dis}_{t, s, i}) h^{\rm req} \le \overline{e}_{t, i}.
\end{equation}
where $h^{\rm req}$ is the required energy reserve ratio, $\theta_i$ is the self-discharge coefficient of the energy storage units, $\eta^{\rm ch}_{i}$ and $\eta^{\rm dis}_{i}$ are the charging and discharging efficiencies, respectively.

\paragraph{Fast-response constraints} The VPP needs to ensure that the fast-response capability of the resources satisfies the response demand, formulated as ($\forall i, \forall i$):
\begin{subequations}\label{model_ramp}
  \begin{equation}\label{model_ramp1}
    0 \le R^{\rm up}_{t, i} \le \overline{R}^{\rm up}_{t, i}, \ 0 \le R^{\rm dn}_{t, i} \le \overline{R}^{\rm dn}_{t, i}
  \end{equation}
  \begin{equation}\label{model_ramp2_1}
    h^{\rm R} R^{\rm up}_{t, i} \le Cap^{\rm up}_{t, i} + Cap^{\rm dn}_{t, i}
  \end{equation}
  \begin{equation}\label{model_ramp2_2}
    h^{\rm R} R^{\rm dn}_{t, i} \le Cap^{\rm up}_{t, i} + Cap^{\rm dn}_{t, i}
  \end{equation}
  \begin{equation}\label{model_ramp3_1}
    \sum_i R^{\rm up}_{t, i} \ge \varepsilon^{\rm up}_t Cap_t^{\rm reg} + \varepsilon^{\rm res}_t Cap_t^{\rm res}
  \end{equation}
  \begin{equation}\label{model_ramp3_2}
    \sum_i R^{\rm dn}_{t, i} \ge \varepsilon^{\rm dn}_t Cap_t^{\rm reg}
  \end{equation}
\end{subequations}
where $R^{\rm up}_{t, i}$ and $R^{\rm dn}_{t, i}$ are the up- and down-ramp capabilities provided by resource $i$, respectively, subject to its response limits $\overline{R}^{\rm up}_{t, i}$ and $\overline{R}^{\rm dn}_{t, i}$ (\ref{model_ramp1}). In addition to meeting the response requirements, resources also need to be capable of fast response for a certain duration $h^{\rm R}$. Within $h^{\rm R}$, the change in power cannot exceed the adjustable capacity of the resource's power (\ref{model_ramp2_1}-\ref{model_ramp2_2}). In our case study, we set the parameter \(h\) such that the resources can follow the control signal within the response capacity.
$\varepsilon^{\rm up}_t$ and $\varepsilon^{\rm dn}_t$ are given by Equ. (\ref{model_signal_ramp}). $\varepsilon^{\rm res}_t$ is the fast-response requirement for reserve services, which is typically stipulated by the market, such as being able to complete the response within 10 minutes.

Note that while only Equ. (\ref{model_ramp}) explicitly expresses the response constraints, the determination of these constraints also relies on the definition in Equ. (\ref{model_def_capacity}) and the associated constraints in Equ. (\ref{model_capacity}) which are also the contributions of this study.
The power disaggregation model is consistent with the model in \cite{chen_real-time_2024}, but fast-response capabilities of the resources should be considered in the power limits, which is easy to implement and omitted here.

\section{Case Study}\label{sec_case_study}

We used Gurobi (V11.0.0)
and MATLAB (R2023b)
with YALMIP
to solve the optimization problems. The computations were executed on a workstation with an Intel Core i9-10900X CPU (3.7 GHz) and 128 GB of RAM. The complete model, code, and data have been open sourced and are available on \cite{rick10119_github}.

\subsection{Parameters and case settings}

\begin{figure}[!t]
  \centering
  \includegraphics[width=3.0in]{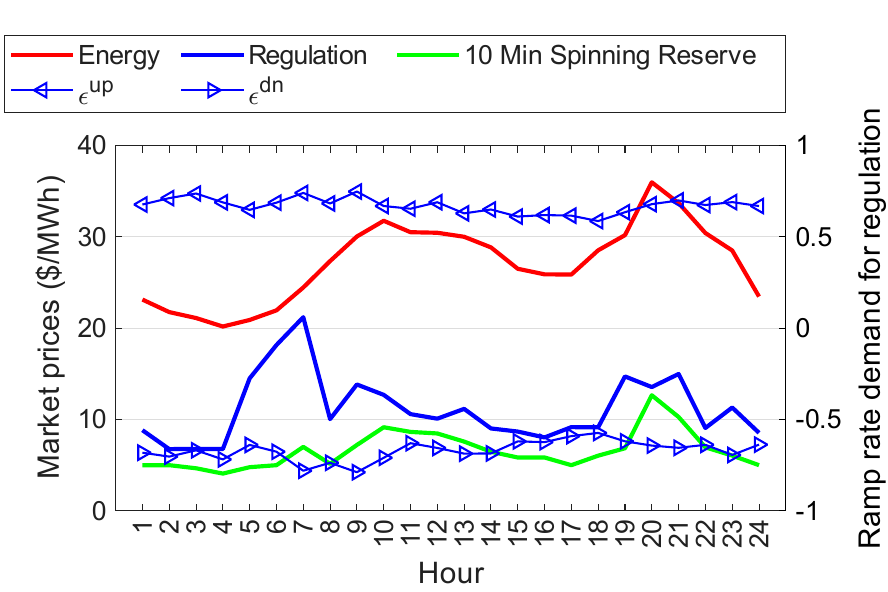}
\caption{Hourly market price and ramp rate demand for regulation services.}
  \label{price_RR_demand}
\end{figure}

We validate the effects of introducing fast-response constraints through a case study involving a small-scale VPP composed of an ES, an EV fleet, and a TCL. The ES power/energy capacity is 0.5 MW/1 MWh, the EV fleet consists of 30 bidirectional 7.68 kW EVs, and the TCL has a rated power of 1 MW. The specific parameters are borrowed from \cite{chen_scheduling_2021, lu_data-driven_2021, li_emission-concerned_2013}. We assume that the power of the ES and EVs allows their power to be adjusted between 0 and the rated power within 5 seconds, while the TCL, due to limitations of the electromechanical components and closed-loop control in air conditioning, has a ramp rate restricted to 100 kW/min, which is close to the value in Ref.~\cite{cui_evaluation_2018}. We use the day-ahead market prices of NYISO on April 13, 2024, as boundary conditions, and the prediction and simulation methods for the frequency regulation signal are consistent with those in Ref.~\cite{chen_real-time_2024}, where historical RegD data from PJM are used since no control signal data for NYSIO is available. When determining the ramp capacity requirements for frequency regulation services through historical regulation signals, the sampling interval is set to 1 min, with a confidence level of 90\%; for the ramp capability requirements of reserve services, according to market requirements, the response needs to be completed within 10 minutes.

Under the above settings, we carried out simulations under four cases:
Case 1: Each resource participates in the market independently rather than forming a VPP collectively. To differentiate, we use the resource names to represent the scenarios. For instance, Case 1-TCL denotes the situation where only the TCL provides ancillary services; no response requirement is considered.
Case 2: Each resource participates in the market independently, considering fast-response requirements.
Case 3: Resources form a VPP to participate in the market, without considering the fast-response requirement.
Case 4: Resources form a VPP to participate in the market, considering the fast-response requirement.

In the simulation, the metrics include: a) Perf. score: The score is calculated using the PJM method for the performance score for regulation services. b) AS profit: The change in total profit after providing ancillary services compared to the profit when only participating in the energy market.

\subsection{Results and comparison}

\begin{figure}[!t]
  \centering
    \includegraphics[width=3.0in]{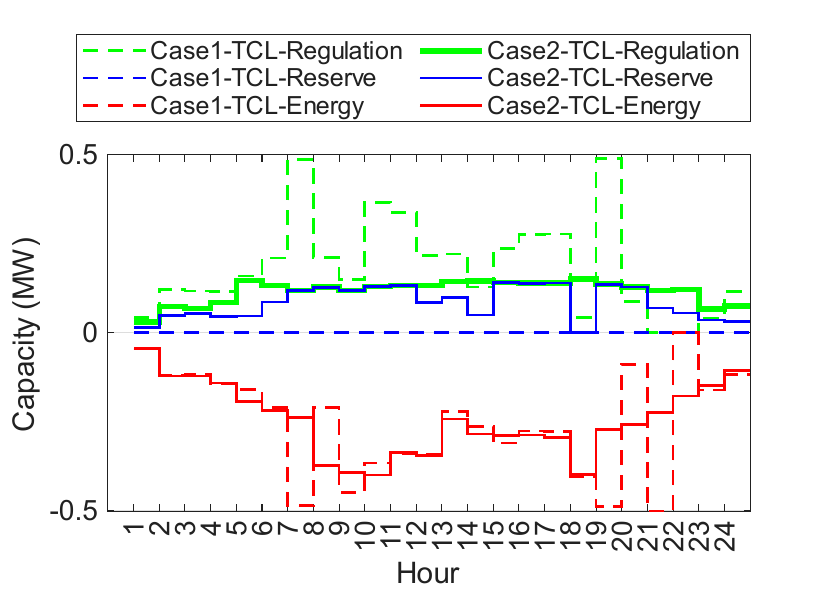}
    \includegraphics[width=3.0in]{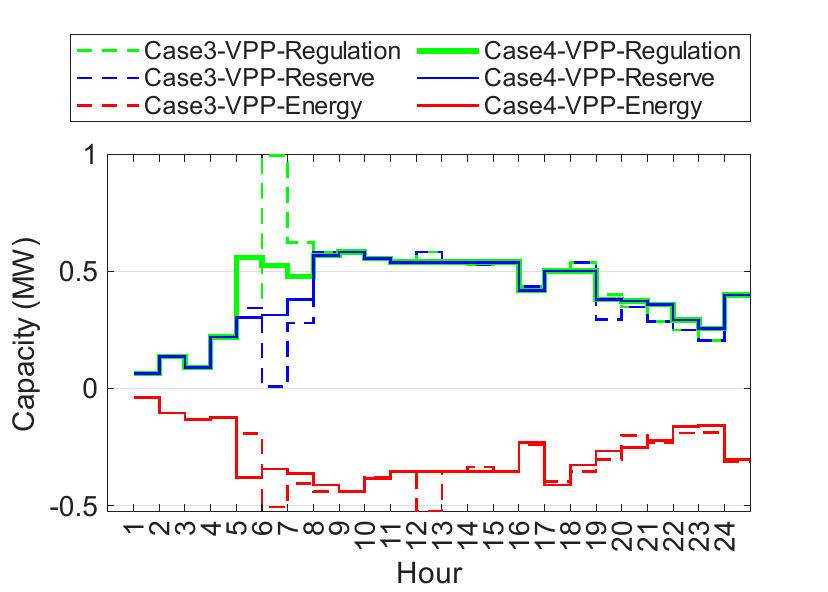}
  \caption{Hourly energy and ancillary service bids in the market. Top: Bids of the TCL in Case 1/Case 2. Bottom: Bids of the VPP in Case 3/Case 4.}
  \label{bid_result}
\end{figure}

\begin{table}[!t]
  \caption{Simulation Results of the Cases}
  \label{tab_c1_c2}
  \centering
  \renewcommand{\arraystretch}{1}
  \begin{tabular}{lrrrrr}
\toprule
Case      & \begin{tabular}[c]{@{}l@{}}Avg. \\  perf.\\ score\end{tabular} & \begin{tabular}[c]{@{}l@{}}Min. \\  perf.\\ score\end{tabular} & \begin{tabular}[c]{@{}l@{}}AS \\ income\\ (\$)\end{tabular} & \begin{tabular}[c]{@{}l@{}}Operation \\ cost \\ (\$)\end{tabular} & \begin{tabular}[c]{@{}l@{}}AS \\ profit\\ (\$)\end{tabular} \\ \midrule
Case1-ES  & 0.997                                                            & 0.997                                                            & 37.4                                                     & 32.2                                                           & 4.0                                                      \\
Case2-ES  & 0.997                                                            & 0.997                                                            & 37.4                                                     & 32.2                                                           & 4.0                                                      \\
Case1-EV  & 0.997                                                            & 0.997                                                            & 26.3                                                     & 6.5                                                            & 17.5                                                     \\
Case2-EV  & 0.997                                                            & 0.997                                                            & 26.3                                                     & 6.5                                                            & 17.5                                                     \\
Case1-TCL & 0.833                                                            & 0.590                                                            & 40.7                                                     & 0.0                                                            & 33.5                                                     \\
Case2-TCL & \textbf{0.960}                                                            & \textbf{0.912}                                                            & \textbf{44.9}                                                     & 0.0                                                            & \textbf{39.2(+17\%)}                                                     \\
Case3-VPP & 0.952                                                            & 0.857                                                            & 176.4                                                    & 65.2                                                           & 98.8                                                     \\
Case4-VPP & \textbf{0.964}                                                            & \textbf{0.921}                                                            & 176.3                                                    & \textbf{59.0(-9\%)}                                                           & \textbf{106.5(+8\%)}                                                    \\ \bottomrule
\end{tabular}
\end{table}

Table~\ref{tab_c1_c2} demonstrates the impact of introducing fast-response constraints on the performance of the VPP in the market relative to the existing VPP operation strategies. First, for the scenarios where the ES and EV individually provide ancillary services (Case 1), the introduction of fast-response constraints (Case 2) does not alter their performance. This indicates that the fast-response constraints we designed do not limit their bidding when the resources have sufficient fast-response capability; hence, they are compatible with existing VPP operation strategies. Second, for scenarios where the TCL individually provides ancillary services (Case 1) and where the resources form a VPP to provide ancillary services (Case 3), the introduction of fast-response constraints (Cases 2 and 4) enhances both the performance score and the profit from the ancillary market.

Fig.~\ref{bid_result} shows the changes in the bidding results when fast-response constraints are introduced. In particular, when the TCL participates in the market alone (Fig.~\ref{bid_result}(a)), the introduction of fast-response constraints (solid lines) leads to a reduction in the capacity bids in the frequency regulation market, which has high fast-response requirements, and an increase in the capacity bids in the reserve market, which has lower fast-response requirements. On the other hand, in the case of joint frequency regulation by multiple resources (Fig.~\ref{bid_result}(b)), the impact of introducing fast-response constraints is less due to the excellent fast-response capability of resources such as ES, which can compensate for the lack of TCL fast-response capability.

\begin{figure}[!t]
  \centering
    \includegraphics[width=3.0in]{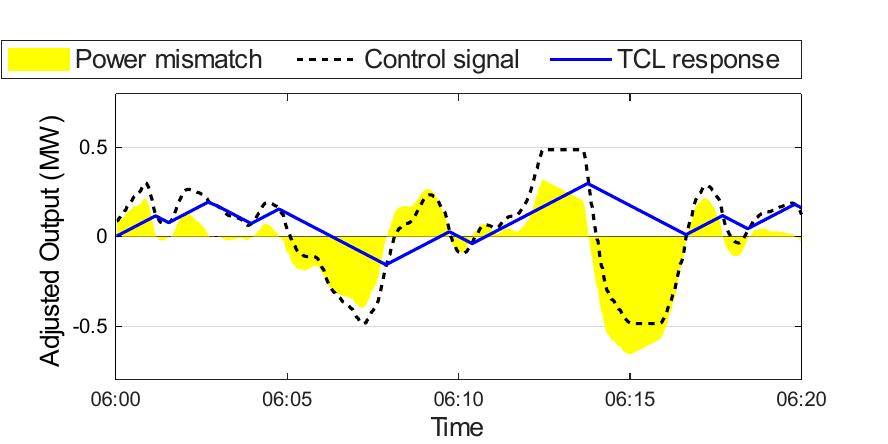}
    \includegraphics[width=3.0in]{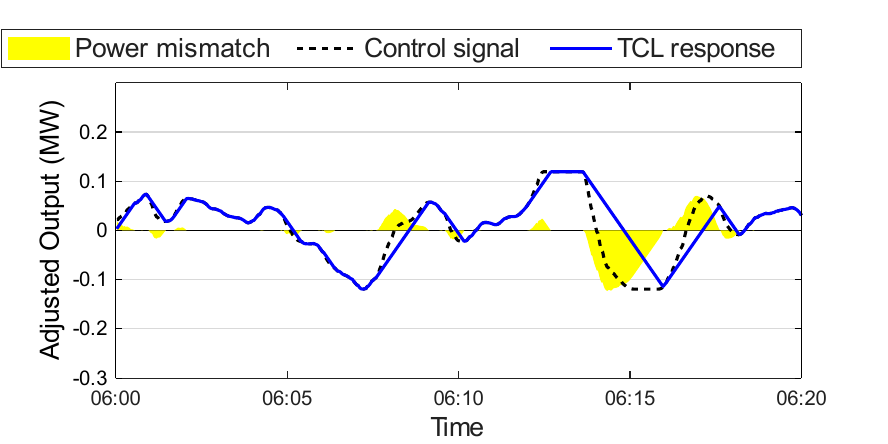}
  \caption{Typical response of the TCL to control signals when participating in the market individually. Top: Case 1-TCL. Bottom: Case 2-TCL.}
  \label{typical_hour}
\end{figure}

Fig.~\ref{typical_hour} shows the response to control signals during typical hours when the TCL alone provides ancillary services. At Hour 6, the frequency regulation market price is relatively high; therefore, under Case 1, the TCL reports a frequency regulation capacity of 0.5 MW. However, its limited fast-response capability prevents it from fully following the rapidly changing control signal (Fig.~\ref{typical_hour}(a)), resulting in its actual response capacity, performance score, and market revenue being lower than expected. The performance score is as low as 0.59 at its lowest point, which does not meet the threshold for participation in the PJM RegD market. In contrast, our method (Case 2) restricts the bids after balancing the response demand and the fast-response capability of the resources. This allows the TCL to follow the ramp of the control signal with an appropriate probability (Fig.~\ref{typical_hour}(b)), thereby obtaining an average performance score of 0.960 and increasing market profits to \$39.2 (up by 17\%).

In Cases 3 and 4, despite the strong fast-response capability of EVs and ESs, our method can still enhance VPP operation by reducing response costs (-9\%) and increasing overall profits (+8\%) by avoiding reporting excessively high capacity. This is because, under Case 3, the response volume exceeding the fast-response capability of the TCL is replaced by resources such as ES, which will generate more costs.

\section{Conclusion}\label{sec_conclusion}

In this paper, we integrate fast-response capability considerations into the operation strategies of VPPs participating in the ancillary services market. 
Specifically, we proposed a practical method to help the VPP balance the fast-response requirements of the ancillary services market with its fast-response capabilities, thereby formulating more reasonable operation strategies. This can improve the operation strategy of the VPP, enhancing its performance score and revenue. 
Our method significantly enhances both the technical and commercial viability of VPPs, thus better supporting the integration of renewable energies.




\section*{Acknowledgments}
Generative AI was used to enhance the grammar and readability of this paper.

\appendices
\section{Complete VPP Operation Model}\label{apd_model}

The VPP aims to maximize its profit $v^{\rm VPP}$ over the time horizon, which is the income from the power market minus operation costs, formulated as:
\begin{equation}\label{model_vpp_profit_total}
  v^{\rm VPP} = \sum_{t=1}^{T}(Income^{\rm e}_{t}
  + Income^{\rm reg}_{t} + Income^{\rm res}_{t} - Cost_{t})
\end{equation}
where $Income^{\rm e}_{t}$, $Income^{\rm reg}_{t}$, and $Income^{\rm res}_{t}$ are the incomes from the energy market, regulation market, and reserve market, respectively. $Cost_{t}$ is the cost of dispatching the DERs. The incomes and costs are calculated as expected values based on the market price forecast and the distribution of control signals. To keep the notation simple, we use $s = (s^{\rm reg}, s^{\rm res})$ to represent the scenario of the control signal; thus, $\pi_{t, s}$ is the probability of the joint distribution of the frequency regulation and reserve signals. $\delta^{\rm reg}_s$ and $\delta^{\rm res}_s$ are the representative values of regulation and reserve signals to represent scenario $s$, respectively, e.g., $\delta^{\rm reg}_s = 0.95$ for the scenario of $\delta^{\rm reg} \in [0.9, 1)$. The detailed calculation of the profit model is as follows:
\begin{subequations}\label{model_vpp_profit}
  \begin{equation}\label{model_vpp_profit_e}
    Income^{\rm e}_{t} =  Pr^{\rm e}_{t} (\tilde{p}_t + \underset{s \in S}\sum \pi_{t, s} (\delta^{\rm reg}_s Cap^{\rm reg}_t + \delta^{\rm res}_s Cap^{\rm res}_t)) \Delta t
  \end{equation}
  \begin{equation}\label{model_vpp_profit_reg}
    Income^{\rm reg}_{t} =  s^{\rm perf} (Pr^{\rm reg, cap}_{t} + Pr^{\rm reg, mil}_{t} a^{\rm mil}_{t}) Cap^{\rm reg}_t  \Delta t
  \end{equation}
  \begin{equation}\label{model_vpp_profit_res}
    Income^{\rm res}_{t} =  Pr^{\rm res}_{t} Cap^{\rm res}_t   \Delta t
  \end{equation}
  \begin{equation}\label{model_vpp_profit_cost}
    Cost_{t} = \underset{i}{\sum} \underset{s \in S}{\sum}\pi_{t, s} (c^{\rm dis}_{i} p^{\rm dis}_{t, s, i} + c^{\rm ch}_{i} p^{\rm ch}_{t, s, i}) \Delta t
  \end{equation}
\end{subequations}
where $Pr^{\rm e}_{t}$, $Pr^{\rm reg, cap}_{t}$, $Pr^{\rm reg, mil}_{t}$, and $Pr^{\rm res}_{t}$ are the prices of energy, regulation capacity, regulation mileage, and reserve service, respectively. $\tilde{p}_t$ is the amount of energy purchased by the VPP in the market, which is also the baseline output of the VPP. Following the convention of power plants, $p > 0$ signifies the injection of energy into the power grid. $Cap^{\rm reg}_t$ and $Cap^{\rm res}_t$ are the bids for regulation and reserve capacities, respectively. The exchanged energy with the grid (\ref{model_vpp_profit_e}), i.e., the sum of the baseline output and the deployed energy for regulation and reserve, are settled at energy prices. The income from the regulation market is performance-based (\ref{model_vpp_profit_reg}), where the performance score $s^{\rm perf}$ and the expected regulation mileage $a^{\rm mil}_{t}$ are regarded as known parameters given by the GO. $c^{\rm dis}_{i}$ and $c^{\rm ch}_{i}$ (\$/kWh) are the levelized operation cost of discharging and charging the DERs (\ref{model_vpp_profit_cost}), respectively, here regarded as parameters submitted by the DERs. $p^{\rm dis}_{t, s, i}$ and $p^{\rm ch}_{t, s, i}$ are the discharging and charging power of DER $i$ at time interval $t$ and signal scenario $s$.

Apart from the constraints in Section~\ref{sec_model_operation}, the VPP needs to ensure that the operation of each DER $i$ is feasible. The constraints are formulated in the form of generalized ES models, which can express the technical constraints of typical DERs such as ES, EV, TCL, and photovoltaic (PV), including the power limits (\ref{model_std_powerLimit}), the state limits (\ref{model_std_energyLimit}), the energy balance equation (\ref{model_std_energyChange}), and the initial states (\ref{model_std_energyInit}):
\begin{subequations}\label{model_std}
  \begin{equation}\label{model_std_powerLimit}
    \underline{p}^{\rm dis}_{t, i} \le p^{\rm dis}_{t, s, i} \le \overline{p}^{\rm dis}_{t, i}, \ \underline{p}^{\rm ch}_{t, i} \le p^{\rm ch}_{t, s, i} \le \overline{p}^{\rm ch}_{t, i}, \ \forall  t, \forall s
  \end{equation}
  \begin{equation}\label{model_std_energyLimit}
    \underline{e}_{t, i} \le e_{t, i} \le \overline{e}_{t, i},\ \forall t
  \end{equation}\vspace{-3.8ex}
  \begin{align}\label{model_std_energyChange}
    e_{t+1, i} = \theta_i e_{t, i} +
    \underset{s}{\sum}\pi_{t, s} (\eta^{\rm ch}_{i} p^{\rm ch}_{t, s, i} -  \frac{1}{\eta^{\rm dis}_{i}} p^{\rm dis}_{t, s, i}) \Delta t  + w_{t, i} \Delta t, \ \forall t
  \end{align}\vspace{-3.8ex}
  \begin{equation}\label{model_std_energyInit}
    e^{\rm init}_{i} - e_{t, i} = 0,\ t = 1
  \end{equation}
\end{subequations}
where $\theta_i$ is the self-discharge coefficient of the energy storage units, $\eta^{\rm ch}_{i}$ and $\eta^{\rm dis}_{i}$ are the charging and discharging efficiencies, respectively, and $w_{t, i}$ is the net energy consumption of DER $i$ at time interval $t$.
The heat dissipation and impact of the ambient temperature on the room temperature are represented by the dissipation rates $\theta$ and $w$, respectively, of the TCLs.

\ifCLASSOPTIONcaptionsoff
  \newpage
\fi



\bibliographystyle{IEEEtran}
\bibliography{reference}
\end{document}